\documentclass{svproc}
\usepackage{graphicx}
\usepackage{mathtools}

\newcommand{\mean}[1]{\left\langle{#1}\right\rangle}
\newcommand{\condl}{\mean{\bar q q}_l}
\newcommand{\conds}{\langle \bar s s \rangle}
\usepackage{url}

\begin{document}
\mainmatter              % start of a contribution
\title{Patterns and partners within the QCD phase diagram including strangeness}
\titlerunning{Patterns and Partners}  % abbreviated title (for running head)
\author{A. G\'omez Nicola\inst{1} \and J. Ruiz de Elvira\inst{2}
 \and A. Vioque-Rodr\'iguez\inst{1}}
\authorrunning{A. G\'omez Nicola et al.} % abbreviated author list (for running head)
\tocauthor{J. Ruiz de Elvira and A. Vioque-Rodriguez}
\institute{Departamento de F\'{\i}sica
Te\'orica and IPARCOS. Univ. Complutense. 28040 Madrid, Spain,\\
\email{gomez@ucm.es, avioque@ucm.es}
\and
Albert Einstein Center for Fundamental Physics, Institute for Theoretical Physics,
University of Bern, Sidlerstrasse 5, CH--3012 Bern, Switzerland,\\
\email{elvira@itp.unibe.ch}}

\maketitle              % typeset the title of the contribution

\begin{abstract}

We review the current situation of the  pattern of chiral symmetry restoration. In particular, we analyze partner degeneration for $O(4)$ and $U(1)_A$  symmetries within the context of Ward Identities and Effective Theories. The application of Ward Identities to the thermal scaling of screening masses is also discussed. We present relevant observables for which an Effective Theory description in terms of Chiral Perturbation Theory and its unitarized extension are compatible with lattice data even around the transition region.  We pay special attention to the role of strangeness in this context. 
% We would like to encourage you to list your keywords within
% the abstract section using the \keywords{...} command.
\keywords{QCD phase diagram, chiral symmetry restoration.}
\end{abstract}
\section{Introduction}

Over recent years, we are progressively reaching a deeper understanding of the QCD phase diagram and its main properties.  Combined efforts from experiment, lattice simulations  and phenomenology are allowing to access  regions of the $(T,\mu_B)$ plane increasingly richer in baryon density. In particular, beam energy scans \cite{Adamczyk:2017iwn} would reveal whether a critical point exists and the behaviour of QCD matter around it.  This is actually one of the main objectives of the current program of hot and dense QCD matter in lattice and heavy-ion collisions \cite{Ratti:2018ksb,Bazavov:2019lgz}.

In this context, a significative advance has been to realize that the phase boundary lies close to the chemical freeze-out for physical conditions of net baryon number $B$, electric charge $Q$ and strangeness $S$, accesible to experimental heavy-ion experiments. Thus, using hadron statistical models \cite{Andronic:2017pug}, which have been very successful in the past for this purpose, one can fit hadron yields from ALICE data. The result of such fits are points on the freeze-out $(T,\mu_B)$ curve which turn out to overlap with the critical line obtained from lattice collaborations where $\mu_B$ is treated within Taylor expansions to avoid the so-called sign problem \cite{Bazavov:2018mes}. In addition, the study of fluctuations of those very same conserved changes opens up interesting possibilities.  A particularly interesting analysis in this context  regarding strangeness is the study of  crossed susceptibilities  performed in lattice works \cite{Bazavov:2014xya}. This is  relevant because a combination of $BS$ and $QS$ crossed susceptibilities provides a relation between chemical potentials $\mu_{B,S,Q}$. Such relation  can also be  tested at freeze-out with experimental hadron yields fits or with theoretical models such as the Hadron Resonance Gas (HRG).

The $\mu_B=0$ regime is in principle much better understood. Regarding the transition, the most analyzed signals have been  the inflection point of the (subtracted) light quark condensate $\condl=\langle \bar u u + \bar d d \rangle$ and the peak of the scalar (or chiral) susceptibility $\chi_S$. Both reveal a crossover-like transition in the physical  case  ($N_f=2+1$ light flavors and physical quark masses) at $T_c\simeq$ 156 MeV \cite{Bazavov:2018mes,Aoki:2009sc}  which in the chiral limit reduces to $T_c^0\simeq$ 132 MeV \cite{Ding:2019prx} and  becomes a true phase transition, most likely of second order, for two massless flavours \cite{Pisarski:1983ms}.

An open problem in this context is to determine not only the order but the universality class (pattern) of the chiral phase transition. This depends crucially on whether the $U(1)_A$ anomalous symmetry is sufficiently restored at $T_c$ \cite{Pisarski:1983ms,Pelissetto:2013hqa,Brandt:2019ksy}, which may even affect the properties of the possible critical point at $\mu_B\neq 0$ \cite{Mitter:2013fxa}. A second-order $O(4)\equiv SU(2)\times SU(2)$ transition would be preferred in a scenario with  $U(1)_A$ breaking at $T_c$, while a second-order $U(2)\times U(2)$ one would correspond to a $U(1)_A$ restored situation. The latter  may even degenerate into a first order transition for strong enough $U(1)_A$ restoration \cite{Brandt:2019ksy}. 

A useful perspective to explore this problem is the analysis of partners, i.e., hadronic states which should become degenerate  under those symmetries. Consider for instance  the pseudoscalar and scalar nonets  $\pi^a=i\bar\psi_l\gamma_5\tau^a\psi_l$, $\delta^a=\bar\psi_l \tau^a \psi_l$ for isospin $I=1$, $\eta_l=i\bar\psi_l\gamma_5 \psi_l$, $\eta_s=i\bar s \gamma_5 s$, $\sigma_l=\bar\psi_l \psi_l$, $\sigma_s=\bar s  s$ for $I=0$, $K^a=i\bar\psi  \gamma_5 \lambda^a \psi$, $\kappa^a=i\bar\psi  \lambda^a \psi$ $(a=4,5,6,7)$ for $I=1/2$. Here,  $\psi_l$  is the light quark doublet and those states correspond respectively to  the quantum numbers of the pion, $a_0(980)$, light and strange component of the $\eta/\eta'$, light and strange components of the $f_0(500)/f_0(980)$, kaon and $K(800)$ (or $\kappa$). For the isospin $I=0,1$ sector, chiral and $U(1)_A$ transformations connect the bilinears 

\begin{eqnarray}
\pi^a\,&\xleftrightarrow{SU_A(2)}&\sigma, \quad \delta^a\xleftrightarrow{SU_A(2)}\eta_l, \\ \pi^a&\xleftrightarrow{U(1)_A}& \delta^a,  \quad \sigma\xleftrightarrow{U(1)_A}\eta_l
,
\end{eqnarray}
which are the partners that have been studied in recent lattice and theoretical works on this subject. The lattice results are not fully conclusive. On the one hand, for $N_f=2+1$ flavors and physical quark masses, the analysis of~\cite{Buchoff:2013nra} shows degeneracy of $U(1)_A$ partners well above the $O(4)$ ones. On the other hand, $N_f=2$ works~\cite{Aoki:2012yj,Cossu:2013uua,Tomiya:2016jwr,Brandt:2016daq} point to $U(1)_A$ restoration at $T_c$ in the chiral limit, while for massive quarks in those works the strength of $U(1)_A$ breaking increases with the volume  \cite{Brandt:2019ksy}.

\section{Ward Identities}

We have recently analyzed the chiral pattern commented above, exploiting Ward Identities derived formally from the QCD generating functional \cite{GomezNicola:2017bhm,Nicola:2018vug}. In particular, the following identity connects susceptibilities (two-point correlators at $p=0$) in the pseudoscalar $\eta_l$, $\pi$ and crossed $\eta_l\eta_s$ channels with the topological susceptibility of the anomaly operator $A(x)=\frac{3g^2}{16\pi^2}\mbox{Tr}_c G_{\mu\nu}\tilde G^{\mu\nu}$:

\begin{equation}
\chi_P^{ls}(T)=-2\frac{\hat m}{m_s} \chi_{5,disc}(T)=-\frac{2}{\hat m m_s}\chi_{top}(T),
\label{wils5}
\end{equation}
where 
$\chi_{5,disc}(T)=\frac{1}{4}\left[\chi_P^\pi(T)-\chi_P^{ll}(T)\right]$ and $\hat m=m_u=m_d$.  Now, one can choose a $SU(2)_A$ transformation so that

\begin{equation}
\eta_l(x)\xrightarrow{SU_A(2)} -\delta^b (x)\Rightarrow \chi_P^{ls}\xrightarrow{SU_A(2)} 0 , 
\label{chilsvanishing2}
\end{equation}
since $\eta_s$  is invariant under $SU(2)_A$ transformations and the  $\delta\eta_s$ correlator vanishes by parity.   Therefore, from~\eqref{wils5}, the conclusion is that for exact chiral restoration,  where $\delta$ and $\eta_l$ should degenerate, $\chi_{5,disc}$ should vanish as well. Thus, $\pi^a-\eta$ degenerate and the $O(4)\times U(1)_A$ pattern is realized. This should be then the scenario in the chiral limit for two massless flavours at $T_c$, consistently with the lattice results in ~\cite{Aoki:2012yj,Cossu:2013uua,Tomiya:2016jwr,Brandt:2016daq,Brandt:2019ksy}.  For $N_f=2+1$ flavours and physical masses, the strangeness contribution and the large uncertainties for $\delta-\eta_l$ degeneration~\cite{Buchoff:2013nra} might explain a stronger $U(1)_A$ breaking, consistently also with the chiral limit analysis of that collaboration \cite{Bazavov:2018mes}. 

An interesting application of WI in this context is related to the temperature dependence of lattice spatial screening masses $M_i$ ~\cite{Nicola:2018vug,Nicola:2016jlj} for different $i$ channels.  Assuming a scaling $M_i(T)/M_i(0)\sim \left[\chi_i (T)/\chi_i(0)\right]^{-1/2}$, the WI allow to connect $M_i(T)$ with suitably subtracted quark condensates, well under control in lattice simulations. This assumption implies that the zero momentum propagator given by the susceptibilities $\chi_i$  dominates the thermal dependence. One can actually test such scaling laws directly for lattice collaborations  providing data  on both screening masses and quark condensates for the same lattice setup. Such  test has been performed in \cite{Nicola:2018vug} for the $\pi,K,\bar s s$ and $\kappa$ channels, which according to the WI scale as the inverse square root of $\condl$, $\condl+2\conds$, $\conds$ and $\condl-2\conds$ respectively.  The agreement is quite good, with only two fit parameters related to the definition of subtracted condensates. It explains   also the qualitative behaviour of the $M_i(T)$ around $T_c$, from the expected one of the quark condensates involved. Thus, for instance, the rapid growth of $M_\pi (T)$ would be explained by the inverse dependence  $\left[\condl (T)\right]^{-1/2}$ while $M_K$ and $M_{\bar s s}$ are softened by the $\conds(T)$ component. 

%
%
%\begin{figure}
%\centerline{\includegraphics[width=8cm]{smfourchan.pdf}}
%\caption{Comparison of pseudoscalar screening mass ratios and subtracted condensates for the four channels $\pi$, $K$, $\bar s s$ and $\kappa$.  The  lattice data are taken from 
%~\cite{Cheng:2007jq} (condensates) and~\cite{Cheng:2010fe}  (masses).}
%\label{fig:corrfour}
%\end{figure}
%

\section{Effective Theories}

\begin{figure}
\centerline{\includegraphics[width=6.5cm]{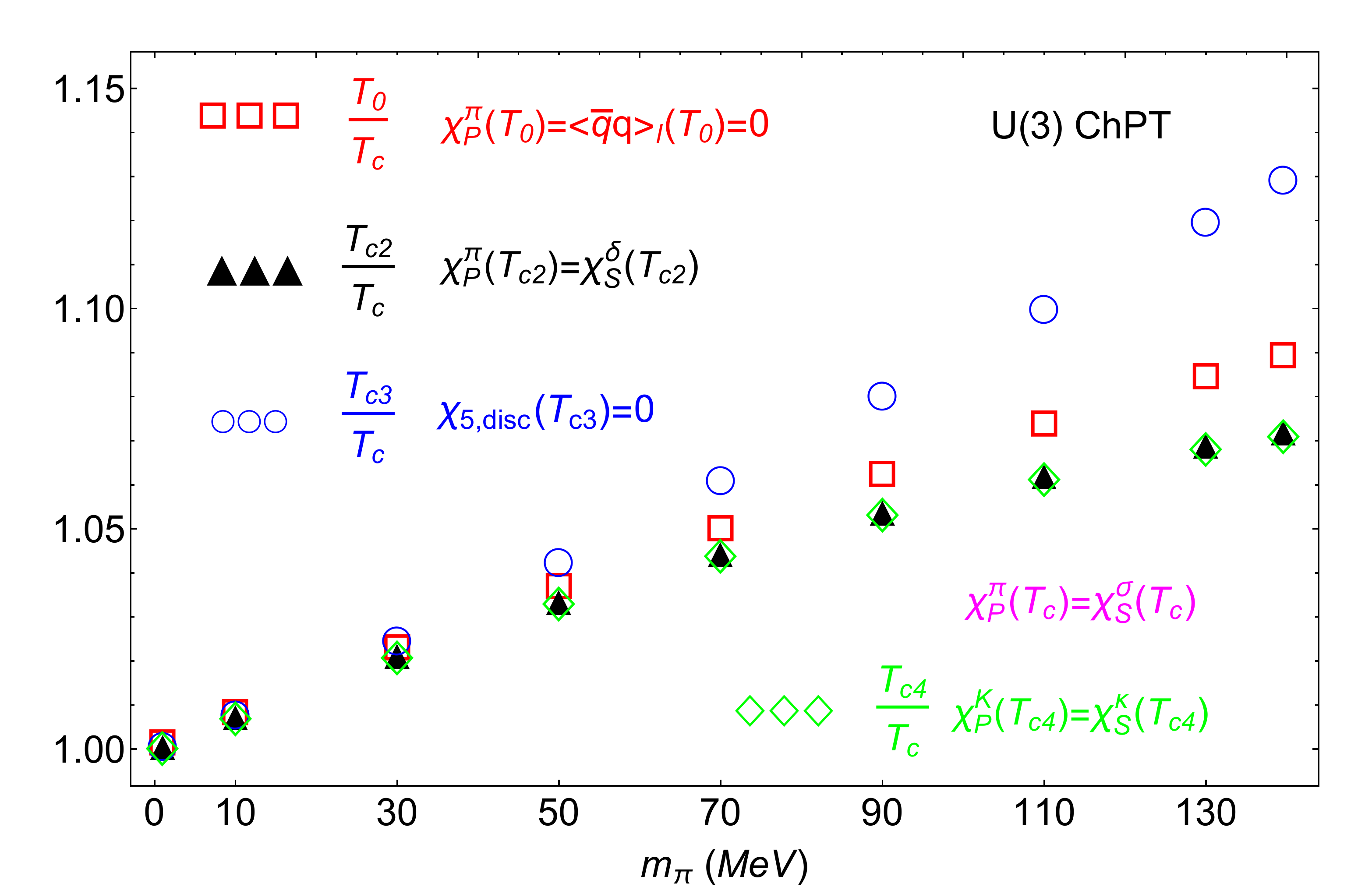}\includegraphics[width=6.5cm]{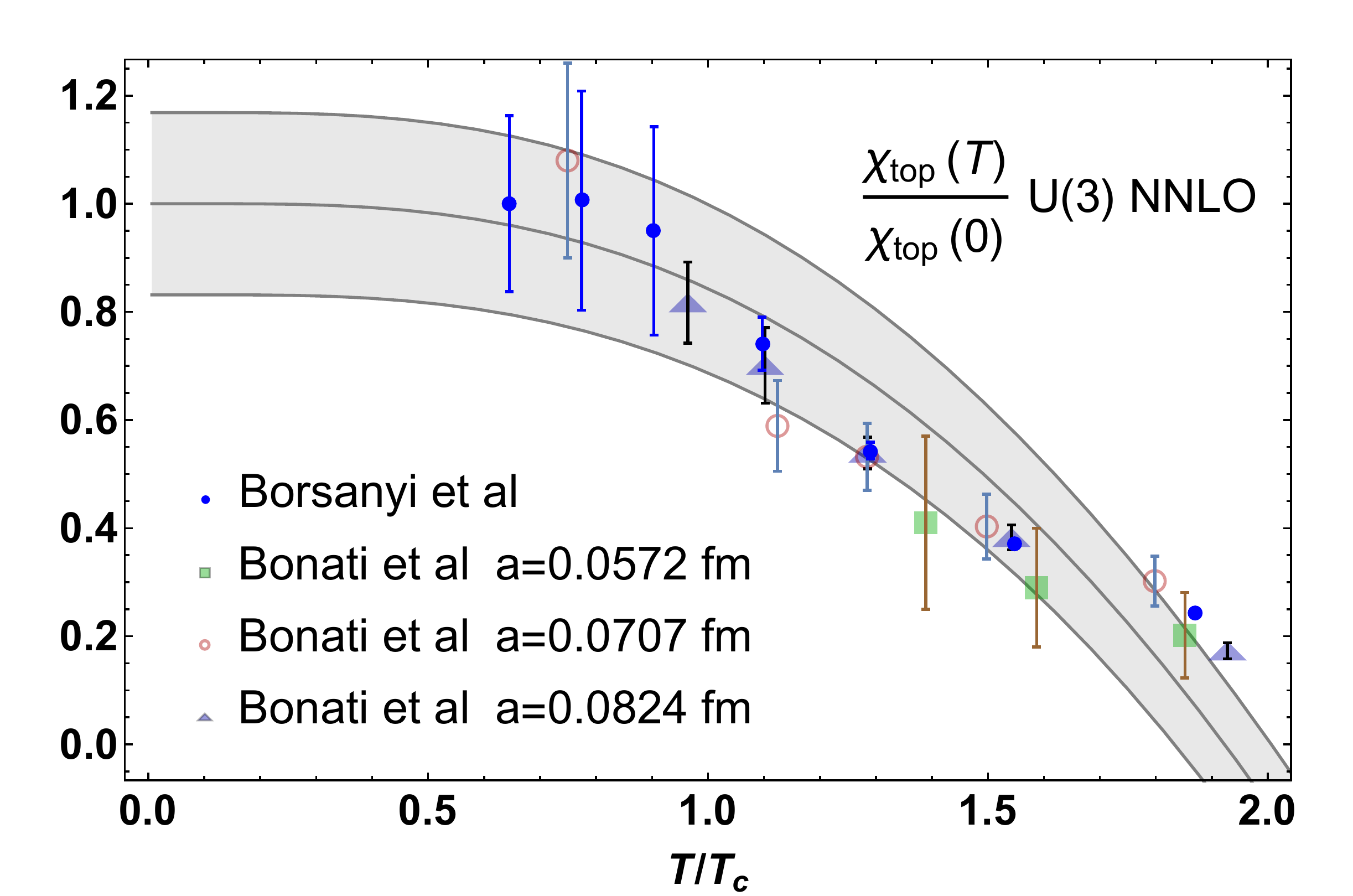}}
\caption{Left: Evolution towards the chiral limit of the different $O(4)$ and $U(1)_A$ restoration temperatures within $U(3)$ ChPT. Right: Temperature dependence of the topological susceptibility calculated within the U(3) formalism compared to lattice data from~\cite{Bonati:2015vqz} and~\cite{Borsanyi:2016ksw} with $T_c=155$ MeV.}
\label{fig:u3chpt}
\end{figure}

Hadronic effective approaches like the HRG or ChPT (for the lightest states) are needed to provide a physically meaningful description below the transition. In connection with our previous discussion, it is worth mentioning that recent analysis within $U(3)$ ChPT (where $N_c^{-1}$ is included in the standard chiral power counting)  have allowed on the one hand to verify the previously mentioned WI \cite{Nicola:2016jlj} and on the other hand to confirm the pattern of $O(4)\times U(1)_A$ restoration in the chiral limit \cite{Nicola:2018vug}. The latter is showed in Fig. \ref{fig:u3chpt} where  pseudocritical temperatures associated to the degeneracy of different $O(4)$ and $O(4)\times U(1)_A$  partners,  converge as the pion mass vanishes.

The $U(3)$ ChPT framework allows also to obtain a quite accurate description of the topological susceptibility $\chi_{top}$ and its thermal dependence \cite{Nicola:2019ohb}. The leading order yields

\begin{equation}
\chi_{top}^{U(3),LO}= \Sigma \frac{M_0^2 {\bar m}}{M_0^2+6B_0{\bar m}} 
\label{chitoploib}
\end{equation}
with  $\Sigma=B_0F^2$  the single-flavor quark condensate in the chiral limit, $B_0=M_{0\pi^\pm}^2/(m_u+m_d)$, with $M_{0\pi^\pm}$ the tree-level mass of the charged pions, $F$  the pion decay constant in the chiral limit, $M_0$  the anomalous part of the $\eta'$ mass and $\displaystyle\bar m^{-1}=\sum_{i=u,d,s} m_i^{-1}$. Expression \eqref{chitoploib} reproduces the known results for two and three light flavours in the limit $M_0\rightarrow\infty$ \cite{Leutwyler:1992yt} as well as the quenched gluodynamics limit for $m_i\rightarrow\infty$ \cite{Witten:1979vv,Veneziano:1979ec}. The NLO corrections can be found in \cite{Mao:2009sy}, while the NLO and NNLO $U(3)$ results at $T=0$ are given in \cite{Nicola:2019ohb}, including the fourth-order cumulant of the topological charge. The contribution of $\eta'$ loops and $\eta-\eta'$ mixing corrections provided by the $U(3)$ formalism are of the same order as the $K,\eta$ $SU(3)$ ones and are compatible with the lattice results in \cite{Bonati:2015vqz,Bonati:2016tvi}. In addition, the large-$N_c$ behaviour of both quantities arising naturally within this formalism agrees also with  lattice analysis \cite{Bonati:2016tvi}. 

The temperature evolution of the topological susceptibility within the $U(3)$ ChPT analysis, showed in Fig. \ref{fig:u3chpt}, is consistent with lattice data, even far beyond the applicability range of the theory.  Although $\chi_{top}(T)$ scales perturbatively as $\condl(T)$ (actually both quantities are proportional at LO), deviations from this behaviour are expected around the transition. Actually, from the WI in \eqref{wils5} and the WI $\chi^\pi_P=-\condl/\hat m$, an additional contribution proportional to $\chi_P^{ll}(T)$  is present, consistently with the existence of a sizable gap between chiral and $U(1)_A$ restoration. 

%\begin{figure}
%	\centerline{
%		\includegraphics[width=7cm]{chitopfiniteTu3.pdf}}
%	\caption{Temperature dependence of the topological susceptibility calculated within the U(3) formalism compared to lattice data from~\cite{Bonati:2015vqz} and~\cite{Borsanyi:2016ksw} with $T_c=155$ MeV}
%	\label{fig:temp1}
%\end{figure}

%\begin{table}[h!]
%				\centering
%		\begin{tabular}{|l|l|l|l|}
%			\hline
%			\multicolumn{1}{|c|}{$\chi_{top}^{1/4}$ [MeV]} 
%			&
%			\multicolumn{1}{c|}{U(3)}&
%			\multicolumn{1}{c|}{SU(2)}&
%			\multicolumn{1}{c|}{SU(3)}  \\
%			\hline
%			LO& 74(3) & 75(3) & 75(3) \\
%			\hline
%			NLO& 74(3) & 78(3) & 83(2) \\
%			\hline
%			NNLO& 81(2) &\multicolumn{2}{c}{}\\
%			\cline{1-2}
%		\end{tabular}
%                 		\quad
%                \begin{tabular}{|l|l|l|l|}
%			\hline
%			\multicolumn{1}{|c|}{$b_2=\frac{c_4}{12\chi_{top}}$} 
%			&
%			\multicolumn{1}{c|}{U(3)}&
%			\multicolumn{1}{c|}{SU(2)}&
%			\multicolumn{1}{c|}{SU(3)}  \\
%			\hline
%			LO& -0.01737(4) & -0.02083 & -0.01960 \\
%			\hline
%			NLO& -0.018(2) & -0.029(2) &  -0.025(1)\\
%			\hline
%			NNLO& -0.023(2) &\multicolumn{2}{c}{}\\
%			\cline{1-2}
%		\end{tabular}
%	\caption{Topological susceptibility and the $b_2$ coefficient, calculated in SU(2), SU(3) and U(3) ChPT to LO, NLO and NNLO in the isospin limit as given in  \cite{Nicola:2019ohb}.}
%	\label{tab:num}
%	\end{table}

\begin{figure}
\centerline{\includegraphics[width=8cm]{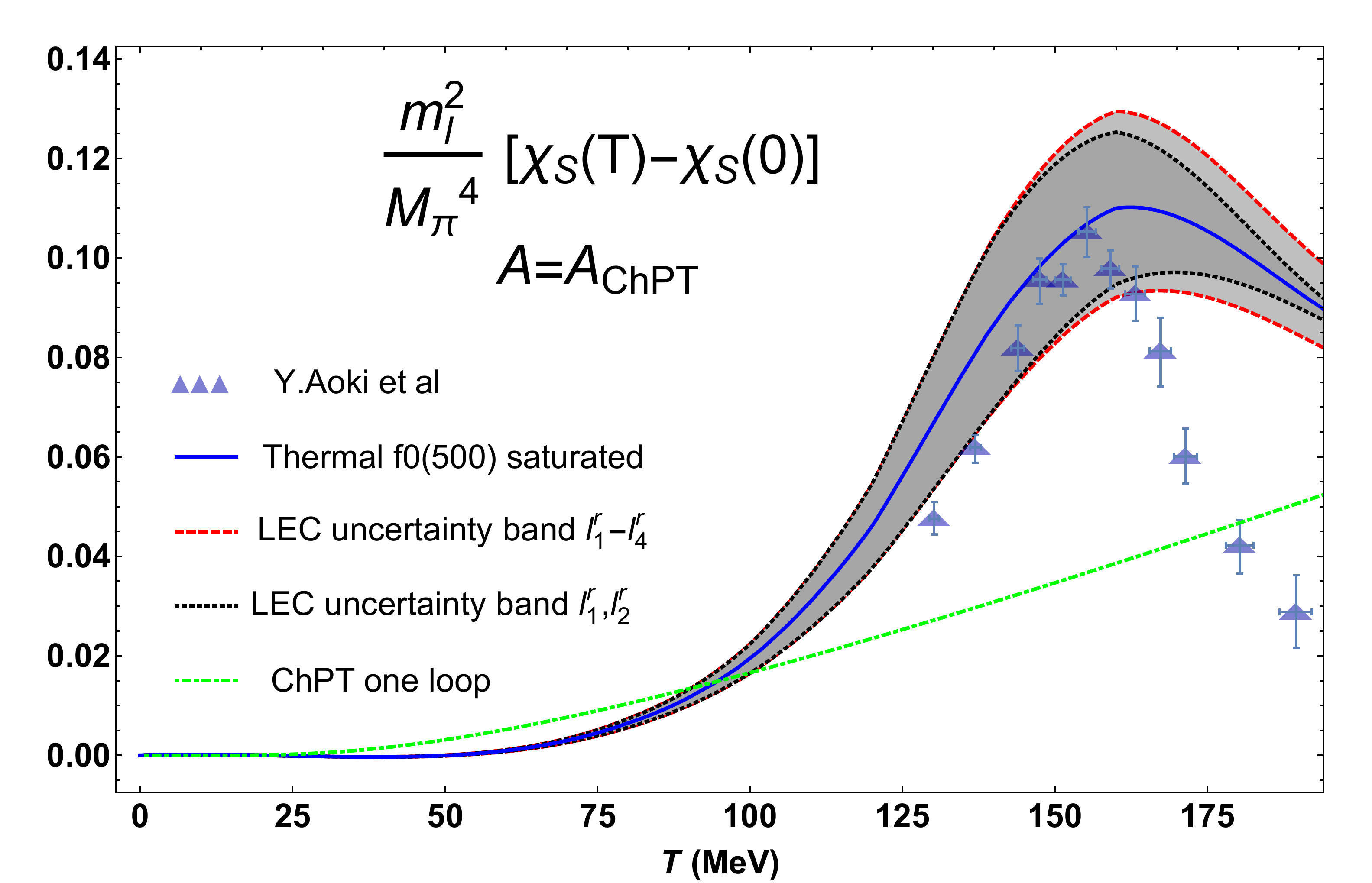}}
\caption{Scalar susceptibility saturated by the unitarized thermal $f_0(500)$ pole, according to \cite{Ferreres-Sole:2018djq}. A normalization factor $A$ has been chosen to match the perturbative ChPT result at $T=0$ and the uncertainty bands given by the low-energy constants (LEC) is shown. Lattice points are taken from \cite{Aoki:2009sc}.}
\label{fig:scalarsus}
\end{figure}

Finally, we remark that combining the standard ChPT expansion with unitarization arguments, one can generate thermal resonances, which show up as second-sheet Riemann poles of meson scattering amplitudes at finite temperature \cite{Dobado:2002xf}. The case of the thermal $f_0(500)$ is particularly important in the present context  since it saturates the scalar susceptibility, giving rise to a peak around the crossover transition compatible with lattice data, as shown in Fig. \ref{fig:scalarsus}, even more accurately than the HRG description \cite{Nicola:2013vma,Ferreres-Sole:2018djq}.

\section{Conclusions}

Despite the recent advances in the understanding of the QCD phase diagram, there are still many relevant open problems such as the nature of the transition, the description of matter rich in baryon density and the critical point. We have showed that the use of theoretical tools such as Ward Identities and Effective Theories allow us to make strong claims about the pattern of the transition.  It  points towards $O(4)\times U(1)_A$ restoration in the  limit of two massless flavours, from the analysis of partner degeneration. Related observables  accurately described within this framework are screening masses, the topological charge distribution and the scalar susceptibility through thermal unitarity. 

\paragraph{Acknowledgements}

Work partially supported by  research contract FPA2016-75654-C2-2-P  (spanish ``Ministerio de Econom\'{\i}a y Competitividad") and the Swiss National Science Foundation, project No.\ PZ00P2\_174228. This work has also received funding from the European Union Horizon 2020 research and innovation programme under grant agreement No 824093. A. V-R acknowledges support from a fellowship of the UCM predoctoral program. 


\end{document}